# Transformable plasmonic helix with swinging gold nanoparticles

Andreas Peil,[a,b] Pengfei Zhan,[a,b] Xiaoyang Duan,[a,b] Roman Krahne,[c] Denis Garoli,[c] Luis M. Liz-Marzán,[d,e,f] Na Liu*[a,b]

[a] A. Peil, Dr. P. Zhan, Dr. X. Duan, Prof. Dr. N. Liu
2. Physics Institute
University of Stuttgart
Pfaffenwaldring 57, 70569 Stuttgart, Germany
E-mail: na.liu@pi2.uni-stuttgart.de
[b] A. Peil, Dr. P. Zhan, Dr. X. Duan, Prof. Dr. N. Liu
Max Planck Institute for Solid State Research
Heisenbergstraße 1, 70569 Stuttgart, Germany
[c] Prof. R. Krahne, Dr. D. Garoli
Instituto Italiano di Tecnologia
Via Morego 30, 16163 Genova, Italy
[d] Prof. Dr. L. M. Liz-Marzán
CIC BiomaGUNE
Paseo Miramón 182, 20014 Donostia/San Sebastián, Spain
[e] Prof. Dr. L. M. Liz-Marzán
Biomedical Networking Center, Bioengineering, Biomaterials and Nanomedicine (CIBER-BBN)
Paseo Miramón 182, 20014 Donostia/San Sebastián, Spain
[f] Prof. Dr. L. M. Liz-Marzán
Ikerbasque, Basque Foundation for Science
43009 Bilbao, Spain

Supporting information for this article is given via a link at the end of the document.

**Abstract:** Control over multiple optical elements that can be dynamically rearranged to yield substantial three-dimensional structural transformations is of great importance to realize reconfigurable plasmonic nanoarchitectures with sensitive and distinct optical feedback. In this work, we demonstrate a transformable plasmonic helix system, in which multiple gold nanoparticles (AuNPs) can be directly transported by DNA swingarms to target positions without undergoing consecutive stepwise movements. The swingarms allow for programmable AuNP translocations in large leaps within plasmonic nanoarchitectures, giving rise to tailored circular dichroism spectra. Our work provides an instructive bottom-up solution to building complex dynamic plasmonic systems, which can exhibit prominent optical responses through cooperative rearrangements of the constituent optical elements with high fidelity and programmability.

## Introduction

To sustain cellular functions, a cell needs to transport a variety of cargos within the complex intracellular milieu. This task is mainly carried out by molecular motors that move along filament tracks.[1] As intricate as a cell, cargos are often transported via the cooperation of multiple motors in order to reach long distances. Such a cooperative behaviour is involved in diverse cellular signalling and sensing processes.[2] Despite the grand challenges to fully understand how cells exactly manage to execute all their intelligent functions, construction of artificial nanosystems by taking inspiration from the working principles that cellular components follow, is undoubtedly an intellectually efficient approach.

The DNA origami technique is likely one of the most successful tools to build functional artificial nanosystems.[3] DNA-based walkers[4], rotors[5], assembly lines[6], transport systems[7], and many other constructs have been realized to directly or partially emulate the behaviour of different cellular components. Taking one step further, DNA-based artificial systems with tailored optical properties have also been developed by integrating optically active elements on DNA origami templates.[8] Among different choices, gold nanoparticles (AuNPs) are currently the key players in DNA-assembled nanophotonics, given their superior properties. To name a few, functionalization of AuNPs with DNA is well established and highly reproducible.[9] AuNPs support localized surface plasmon resonances (LSPRs), which are associated with collective oscillations of the conduction electrons.[10] LSPRs can be readily tuned by changing the material, size, shape, and environment of the NP.[11] In addition, AuNPs do not bleach or blink, offering a long characterization time.

In the past decade, there have been significant research endeavours on building dynamic plasmonic nanostructures, especially, chiral plasmonic nanostructures.[12] It is not only due to the rich physics that one can explore, such as plasmonic circular dichroism (CD), plasmon-induced chirality, *etc*.[13], but also lies in the great potential to create advanced optical sensors, which possess 3D structural reconfigurability, inherent properties of DNA, and highly sensitive CD feedback. In particular, DNA origami templates that are used to host AuNPs can be designed to recognize a broad range of target agents, including nucleic acids[14], aptamers[15], ions[16], proteins[17], enzymes[18], and many other molecules with high programmability, addressability, and sequence specificity. Often, the dynamic tunability of DNA origami-templated plasmonic nanostructures is enabled by the reconfiguration of the templates themselves due to the ease of design and operation. However, this scheme allows for few reconfiguration states and most importantly, the optical elements, such as AuNPs are immobilized on origami, missing many advantages that DNA nanotechnology can offer.

In this work, we demonstrate a transformable plasmonic helix, in which multiple AuNPs can programmably move around a shared DNA origami shaft. To enable the translocations of multiple AuNPs in large leaps without detachment, the AuNPs are guided by DNA swingarms[19] extended from the origami to target positions through sequence-specific DNA interactions. The introduction of the swingarms offers direct transportations of AuNPs in space without undergoing consecutive stepwise movements. This is particularly crucial, when multiple, closely-spaced AuNPs are simultaneously in motion. The transformable configurations of the plasmonic helix and the corresponding CD responses stem from the cooperative rearrangements of the AuNPs in three dimensions. The experimental results are in good agreement with theoretical predictions.

## Results and Discussion

Figure 1a illustrates the plasmonic helix, which is a model chiral system with multiple AuNPs arranged in a helical geometry. The DNA origami shaft is a 24-helix bundle (16 nm in diameter, 107 nm in length), which provides overall 16 binding sites distributed in six rows. Each binding site contains a set of DNA footholds. The numbers of binding sites in these six rows from the top to the bottom are 2, 4, 2, 2, 4, and 2 according to the translocation design. In each row, the spacing between the neighbouring binding sites is 8.4 nm. A swingarm containing a set of relatively long DNA strands is also extended from the origami bundle and positioned in between the binding sites for the attachment of a AuNP in each row. AuNPs are densely coated with DNA strands as feet. Consequently, each of the six AuNPs (1-6, 10 nm in diameter) is attached to one of the six swingarms through DNA hybridization (see Figure 1a). This initial state right after AuNP assembly is named as state 0. More specifically, the six AuNPs are classified in three groups according to their track designs: 1 and 4 (golden), 2 and 5 (silver), 3 and 6 (brown). In the same group, the binding site designs are identical along the track. The related structural details are provided in the Experimental Section of the Supplementary Information and Supplementary Figures S1 to S3.

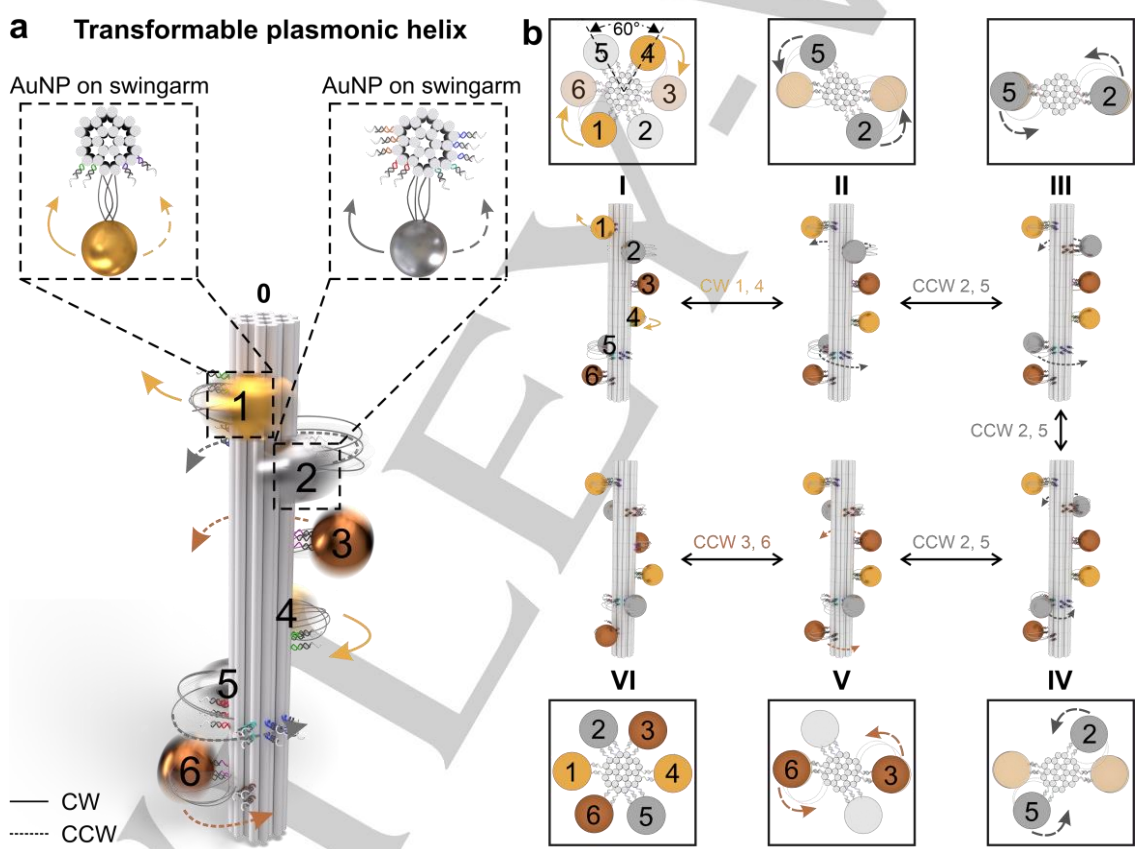

**Figure 1.** (a) Schematic of the transformable plasmonic helix, which comprises a DNA origami bundle and six AuNPs (1 – 6). The six AuNPs attached to six swingarms can move around the bundle along the clockwise (CW, solid arrows) or counterclockwise (CCW, dashed arrows) direction. The AuNPs are classified in three groups according to the track design: AuNPs 1 and 4 (golden), AuNPs 2 and 5 (silver), AuNPs 3 and 6 (brown). (b) Schematic of the six states. The plasmonic helix can be transformed from state I (left-handed, LH) to state VI (right-handed, RH) through cooperative rearrangements of the AuNPs around the origami bundle. The side and top views (black boxes) illustrate the configurations of the AuNPs at different states.

We have defined six states (I-VI), corresponding to six different conformations that the plasmonic helix can reach through cooperative rearrangements of the AuNPs in space (see Figure 1b and Table S2 in the Supplementary Information). State I corresponds to a left-handed (LH) configuration. More specifically, the AuNPs are displaced by 60° with respect to one another (see top view) along the DNA origami bundle to form a LH plasmonic helix. Reversible transitions between neighbouring states are achieved through programmable translocations of the AuNPs by the DNA swingarms around the origami bundle along the clockwise (CW) or counterclockwise (CCW) direction. The AuNP swinging mechanism is explained in detail in Figure 2. As shown in Figure 1b, from state I to state II AuNPs 1 and 4 (golden) are guided to swing CW and subsequently attached to the target binding sites, while the other AuNPs stand still. This process is labelled as CW1,4. By taking successive steps as labelled in Figure 1b, the transformation ends at state VI, which corresponds to a right-handed (RH) configuration.

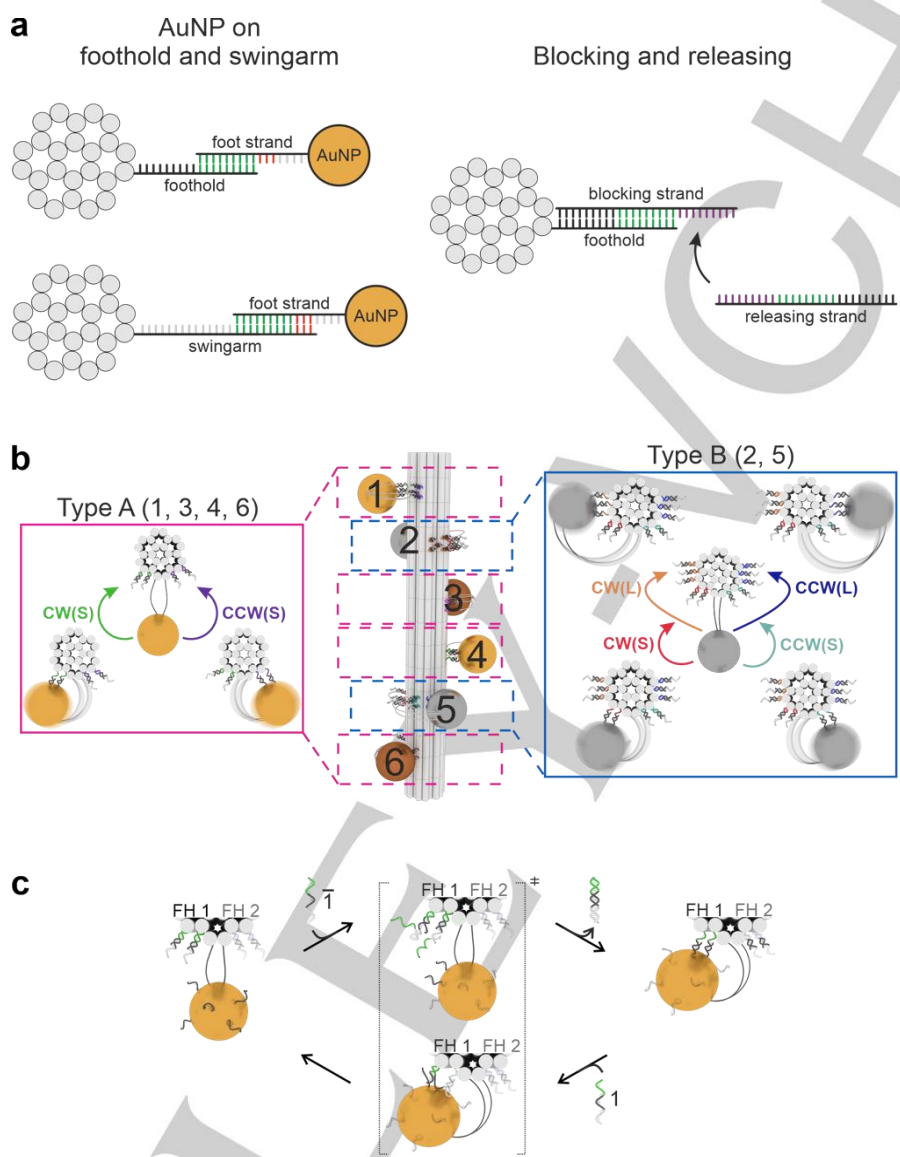

**Figure 2.** (a) Schematic of the AuNP interactions with the swingarm and the DNA footholds. Blocking and releasing strands for deactivation and activation of the foothold are illustrated (for details see the main text). (b) Rows (1, 3, 4, 6) with two binding sites are grouped as type A (pink). Rows (2, 5) with four binding sites are grouped as type B (blue). The AuNP in type A can be guided by the swingarm to reach the left and right binding sites by performing CW(S) and CCW(S) rotations, respectively. S means a small step, 4.2 nm. The AuNP in type B tied to the swingarm, which is longer than that in type A, can also carry out CW(L) and CCW(L) rotations to the outermost left and right binding sites, respectively. L means a large step, 12.6 nm. (c) Working mechanism of the swingarm for the AuNP translocation enabled by toehold-mediated strand displacement reactions. FH1 and FH2 represent any two foothold sites. $\bar{1}$ and 1 represent releasing and blocking strands, respectively.

Figure 2a depicts the scheme of the AuNP-swingarm and AuNP-foothold interactions. The AuNPs are fully functionalized with foot strands. Each swingarm extended from the origami comprises a set of DNA strands. Each strand contains a long single-stranded poly-thymine segment (grey) and a AuNP capturing domain (green and red). Each foothold consists of a unique toehold segment (black) and a AuNP capturing domain (green). Activation and deactivation of the footholds are enabled by toehold-mediated strand displacement reactions upon the additions of releasing and blocking strands (DNA fuels), respectively. As shown in Figure 2a, the blocking strand deactivates the foothold by DNA hybridization. The releasing strand recognizes the toehold region (purple) of the blocking strand and subsequently dissociates it from the foothold. The foothold is thus activated. To avoid nonspecific foot-track interactions, the AuNPs in adjacent rows are functionalized with differently sequenced foot strands (see Supplementary Figure S3 and Supplementary Tables S8 and S9 for the design and sequence details). In other words, AuNPs in alternating rows (1, 3, 5 or 2, 4, 6) are decorated with the same foot strands as feet.

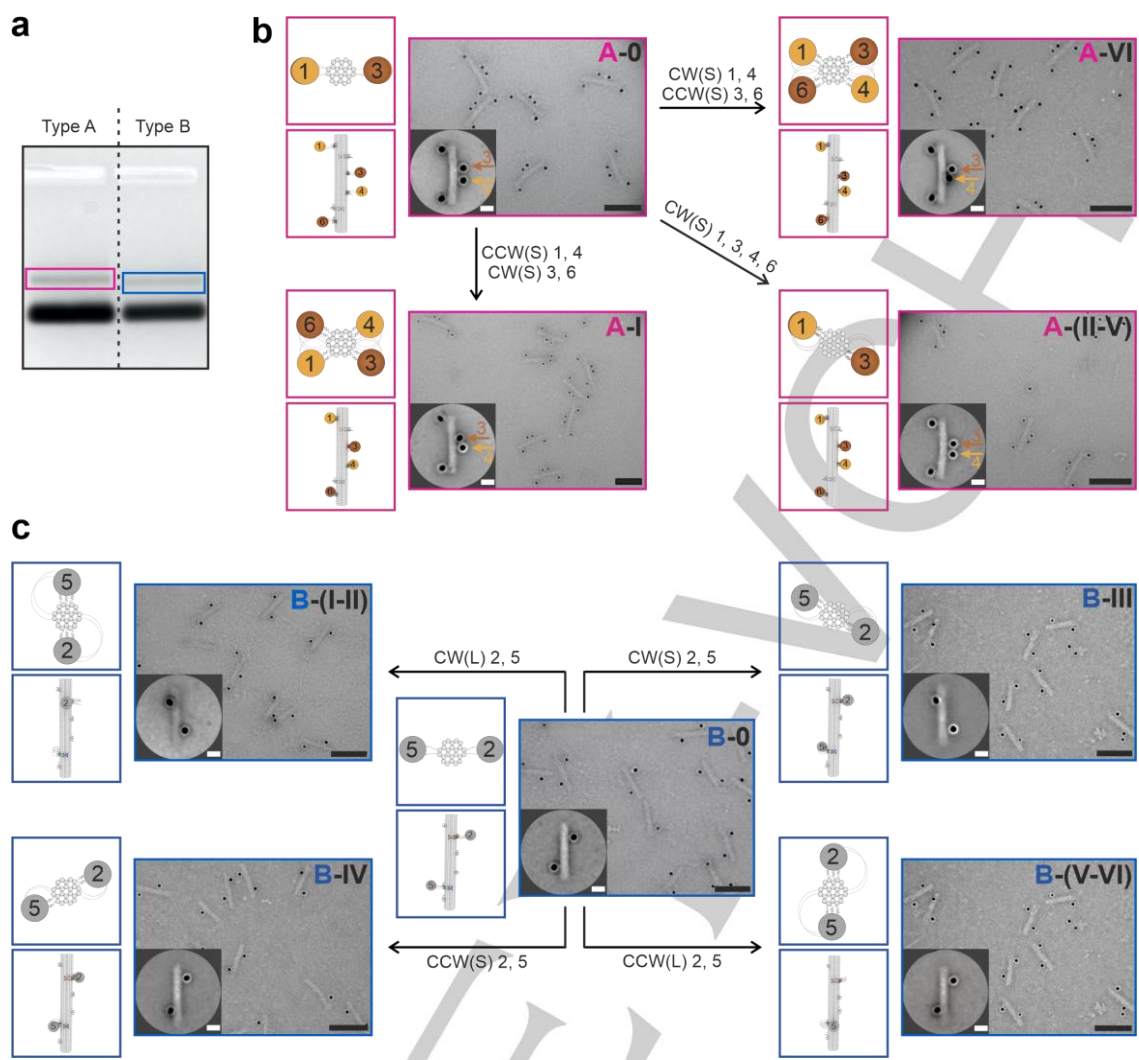

**Figure 3.** (a) Agarose gel electrophoresis image of type A structures comprising AuNPs 1, 3, 4, and 6, as well as type B structures comprising AuNPs 2 and 5. The colour-framed gel bands contain the corresponding structures. (b) Overview and averaged (insets) TEM images of type A structures and the different AuNP rearrangement routes. (c) Overview and averaged (insets) TEM images of type B structures and the different AuNP rearrangement routes. Scale bars, 100 nm. Inset: scale bars, 20 nm.

As shown in Figure 2b, the rows (1, 3, 4, 6) with two binding sites are marked as type A, while the rows (2, 5) with four binding sites are marked as type B. For type A as depicted in the pink frame, the swingarm is positioned in between the two binding sites in each row, so that the AuNP can be swung CW and CCW to the left and right binding sites, respectively, by adding the corresponding DNA fuels. The spacing between the swingarm and the left (right) binding site is approximately 4.2 nm. In type A, the poly-thymine segment in the swingarm is 10 bases. The working mechanism of the swingarm for the AuNP translocation between any two foothold sites (FH1, FH2) is illustrated in Figure 2c. Detailed information of the individual foothold sites can be found in Supplementary Figure S3. Initially the AuNP is tied to the swingarm only. Upon the addition of releasing strands $\bar{1}$, the blocking strands are dissociated from the footholds FH1 through toehold-mediated strand displacement reactions. The AuNP is subsequently bound to the left binding site, carrying out a CW rotation, marked as CW(S). Here, S represents a small step (4.2 nm) to be distinguished from a large step (12.6 nm) in type B (see below). The swinging process is reversible. By adding blocking strands 1, the footholds are deactivated through toehold-mediated strand displacement reactions. The AuNP is thus released from the left binding site but is still fastened to the swingarm. Likewise, the swingarm-tied AuNP can also be directed to the right by binding to the footholds FH2 for a CCW rotation, marked as CCW(S), upon the addition of the corresponding DNA fuels.

For type B, as presented in the blue frame, four binding sites are arranged around the DNA origami bundle in each row. The swingarm is positioned in the middle of these binding sites, two on the left and two on the right. The spacing between the swingarm and the nearer left (right) binding site is ~ 4.2 nm. CW(S) and CCW(S) indicate the CW and CCW rotations of the AuNP to the nearer left and right binding sites, respectively. In order to reach the two outermost binding sites, the swingarm in type B is designed to have a longer poly-thymine segment (14 bases). The spacing between the swingarm and the outermost left (right) binding site is ~ 12.6 nm. The corresponding CW and CCW rotations are marked as CW(L) and CCW(L), respectively. Here, L represents a large step (12.6 nm). Without using the swingarm mechanism for direct transportation, the AuNP would have to carry out consecutive steps by rolling (8.4 nm per step) to reach different states. This not only involves foot-track interactions at each step by adding different DNA fuels, but also the stepwise movements between adjacent binding sites with considerable spacing may easily lead to AuNP detachment during the translocations (see the control experiments in Supplementary Figure S4).

It is worth emphasizing that the DNA swingarms play crucial roles in the self-assembly as well as in the dynamic rearrangements of the AuNPs around the shared DNA origami shaft. First, the swingarms allow for the assembly of the plasmonic helix, comprising multiple, closely-spaced AuNPs with high yield at the initial state, which are crucial for subsequent reconfigurations. Instead, if the AuNPs were only attached to the footholds, detachment of AuNPs may take place already at the initial state. This is because, apart from serving as the binding sites, the footholds also function as the tracks for AuNP movements. Consequently, the AuNP capturing domain of the DNA footholds has to be designed short enough to facilitate the strand displacement reactions. As a result, the assembly yield is in general lower than the case in which all AuNPs are attached to the swingarms with a longer AuNP capturing domain and thus a stronger binding strength (see Figure 2a). Second, because the AuNPs are permanently tied to their respective swingarms during different translocation processes, this greatly helps to prevent AuNP detachment and facilitates foot-track interactions. Third, the sufficient length and flexibility of the swingarms enable the direct transportation of AuNPs to target locations over large distances, avoiding consecutive stepwise movements. Fourth, the swingarms enable reliable structural reconfigurations, especially when multiple, closely-spaced AuNPs are simultaneously in motion.

Type A and type B structures were examined separately using gel electrophoresis and transmission electron microscopy (TEM). As shown by the gel electrophoresis image in Figure 3a, type A structures travel a slightly shorter distance than type B structures due to the difference in the AuNP number, 4 vs. 2. Figure 3b presents the assembly and rearrangement of AuNPs mediated by DNA swingarms in type A structures into different states. At the initial state, labelled as A-0, all the footholds on the origami are deactivated by blocking strands. The AuNPs 1, 3, 4, and 6 are only tied to their respective swingarms. The overview TEM image demonstrates the high-quality of the assembled structures. The class averaged TEM image in the inset reveals more structural information (for details on class averaging see Experimental Section). Due to the long length and flexibility of the swingarms, the AuNPs lie aside to the origami bundle on the TEM grid with noticeable spacing. Departing from state A-0, the four AuNPs are translocated around the origami bundle into different states, following the swinging mechanism presented in Figure 2. To reach state A-I, AuNPs 1 and 4 are swung in a small step CCW (labelled as CCW(S)1,4), while AuNPs 3 and 6 are swung in a small step CW (labelled as CW(S)3,6). The positioning of these four AuNPs is identical to that at state I in Figure 1b and is thus labelled as state A-I. At this state, all four AuNPs are attached to the corresponding footholds, which tighten them towards the origami. The relative spacing changes of the AuNPs to the origami are clearly manifested after the structural transformation. For instance, AuNPs 3 and 4 show an offset with respect to one another, following the helical theme (see arrows in A-I). Following the CW(S)1,3,4,6 route, all four AuNPs carry out small step CW rotations, reaching A-(II-V). Here, II-V means II, III, IV, and V. As shown in Figure 1b, the transitions among states II, III, IV, V are associated only with the translocations of AuNPs 2 and 5 in type B. Therefore, for the type A structure, these states are not distinguishable and can be represented together using A-(II-V). Furthermore, the type A structure can be transformed to state A-VI through CW(S)1,4 and CCW(S)3,6 rotations. It is noteworthy that all states are fully reversible among each other upon addition of the corresponding DNA fuels. Similarly, the assembly and rearrangement of AuNPs by the DNA swingarms in type B structures were studied. The different routes and corresponding TEM images are presented in Figure 3c.

Next, the transformation of the full plasmonic helix comprising both type A and type B AuNPs was investigated (see Figure 4a). At the initial state 0, the six AuNPs tied to their flexible swingarms are lined up at the two sides of the origami bundle in a '1-3-2' pattern on the TEM grid. The plasmonic helix can be switched from state 0 into different states (I-VI) or can be reversibly transformed among all states. The corresponding AuNP translocation routes based on the swingarm mechanism are indicated in the figure. In particular, the structural transformation can be clearly identified, when comparing specific states. For instance, from state 0 to state I, the six AuNPs are changed from a loose '1-3-2' pattern to a more tightly packed helical geometry. It is noteworthy that the AuNPs on the TEM grids can deviate from their equilibrium positions in solution.

To provide further insight into the plasmonic effects derived from the transformable helix, CD spectroscopy was carried out to directly correlate the changing structural conformations with the optical response. Due to the high assembly yield and facile operation of small NPs for translocations, we have utilized 10 nm AuNPs to accomplish the assembly and transformation of the plasmonic helix as shown above. To enhance the weak optical response of 10 nm AuNPs (see Supplementary Figure S9) as well as to reduce the interparticle spacing for stronger coupling effects, AuNP enlargement by electroless deposition of Au in solution was performed prior to CD measurements. TEM images of the structures at different states after Au coating are shown in Figure 4b. The average AuNP diameter was ~18 nm, albeit with some shape inhomogeneity, as expected from electroless deposition. The transformable plasmonic helices with enlarged AuNPs gave rise to meaningful CD spectra at different states (Figure 4c). At state 0, the CD response was close to zero within the wavelength range of interest. This is because the AuNPs tied to the flexible swingarms lead to an undefined structural conformation and thus the CD response is averaged out. At state I, when the AuNPs are tightened and arranged in a helical geometry around the origami bundle by attaching to the corresponding footholds, a bisignate peak-to-dip profile, which is characteristic of LH structures is observed. This observation substantiates the successful AuNP translocations by the swingarms to their target positions, resulting in a plasmonic helix with a well-defined LH configuration. When the plasmonic helix is transformed into states II and III, the CD signals decrease successively. A reverse bisignate dip-to-peak profile starts to develop and becomes clear at state IV. The RH CD response then increases in strength and reaches maximum at state VI. The transformation of plasmonic helices from LH and RH is then completed. Theoretical calculations based on a finite element method were performed to verify the experimental data (for details see Experimental Section). The simulated CD spectra for the different states are shown in Figure 4d, revealing a good agreement with the experimental spectra.

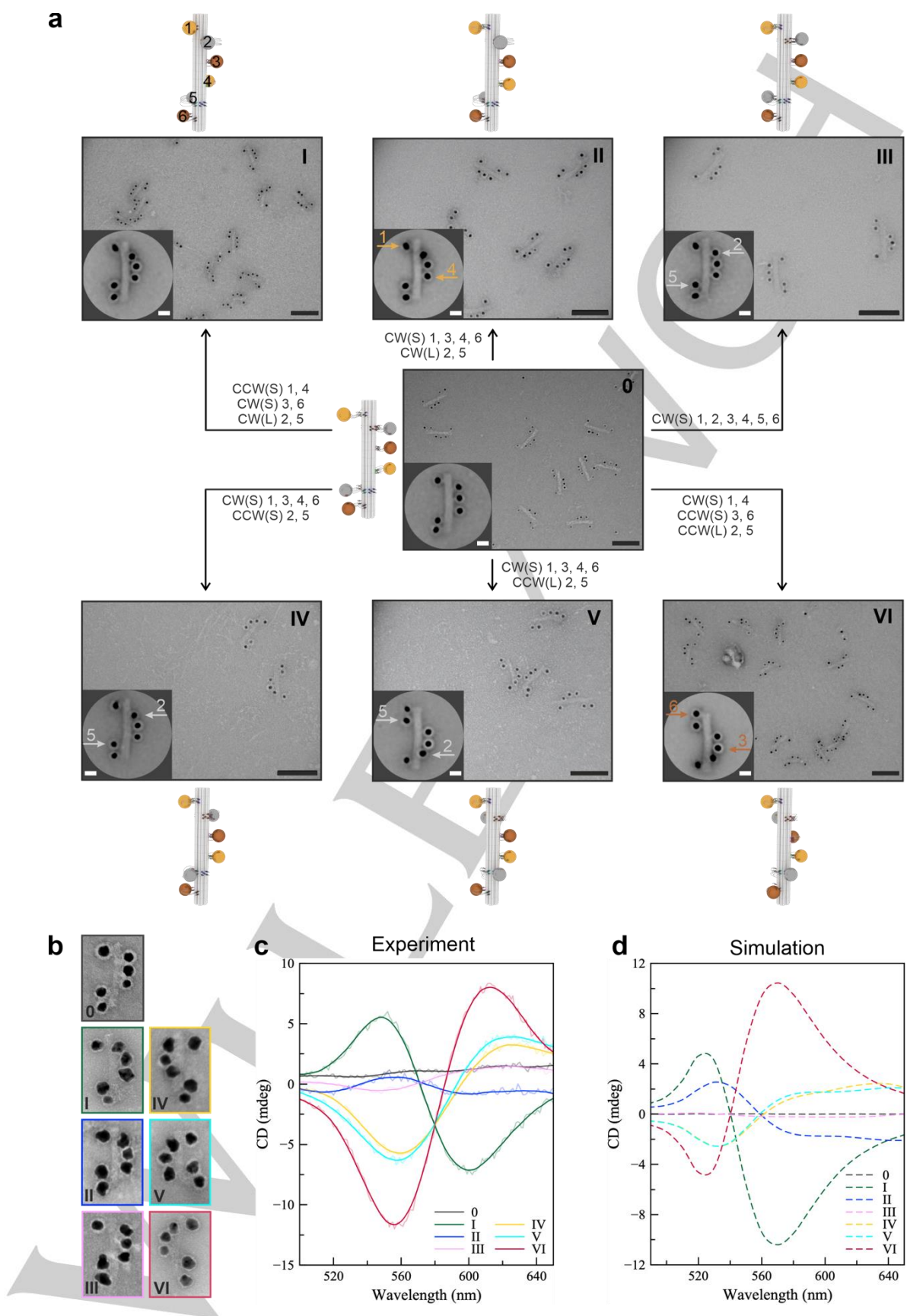

**Figure 4.** Overview and averaged (insets) TEM images of plasmonic helices and different AuNP rearrangement routes. Scale bars, 100 nm. Inset: scale bars, 20 nm. (b) TEM images of representative structures at different states after Au growth. (c) Experimental and (d) calculated CD spectra for different states.

The translocation distance is limited by many design parameters, including the length of the swingarms, the lengths of the foot strands and the footholds, as well as the size of the AuNPs and the positioning of the footholds for binding, among others. With our current design, the largest translocation distance is ~25.2 nm per step (see Figure S11, Supplementary Information). It is possible to enlarge the translocation distance, for instance, by increasing the swingarm length. Finally, we set out to apply the swingarm mechanism in higher-order plasmonic nanoarchitectures. Two versions (A and B) of the origami bundle were created to serve as monomers with differently modified head and tail ends (Figure 5a). The head region (dark green) of monomer B contains single-stranded extensions, which can bind to the tail region (green) of monomer A. The tail region (red) of monomer B is blocked to avoid self-dimerization and enforce the correct relative orientations between monomers A and B for connection. The AuNP-decorated origami dimer structures were purified using agarose gel electrophoresis (Figure 5b). Figure 5c shows representative TEM images of dimer structures at different states and the corresponding AuNP translocation routes. Electroless deposition at different states was also carried out before CD measurements. At state 0, the swingarm-tied AuNPs lie at both sides of the origami dimer bundle on the TEM grid (Figure 5d, 0). Upon AuNP translocations into states I and VI, the LH and RH helical geometries become more visible, respectively. As shown in Figure 5e, the CD spectrum at state 0 exhibited a featureless profile, whereas those at states I (LH) and VI (RH) are nearly mirrored.

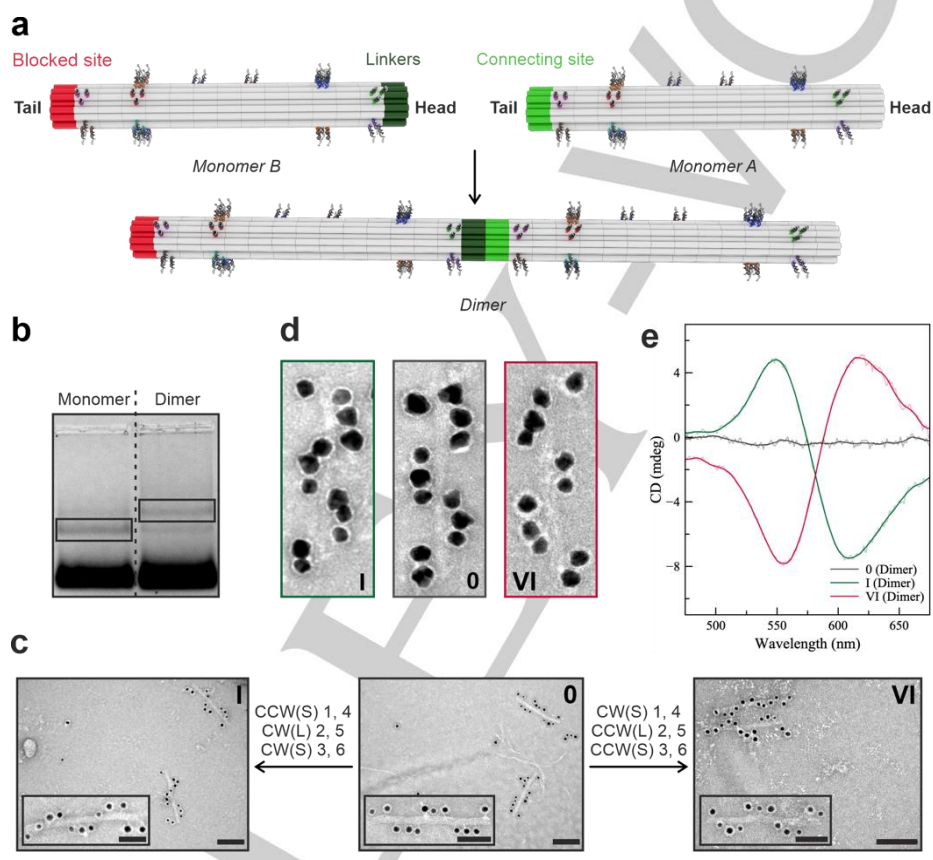

**Figure 5.** (a) Dimerization strategy. Two different origami monomers (A and B) are created with different modifications at the head and tail ends for mutual binding. (b) Agarose gel electrophoresis image of the dimer structures after AuNP assembly. The monomer band is shown as a reference. The framed boxes contain the target structures. (c) Overview and enlarged (insets) TEM images of dimer structures at states 0, I and VI. Scale bars, 100 nm. Inset scale bars, 50 nm. (d) Representative TEM images of the structures after Au coating at states 0, I and VI. The average AuNP diameter is ~18 nm. (e) Experimental CD spectra of the dimer structures after Au coating at states 0, I and VI.

## Conclusion

In conclusion, we have demonstrated a transformable plasmonic helix, in which multiple AuNPs assembled on one origami shaft can be cooperatively rearranged to transit between LH and RH configurations. The swingarm strategy enables direct and reliable AuNP translocations in large steps in a programmable manner. The different states of the system have been characterized by TEM and CD spectroscopy. The experimental results agree well with theoretical predictions. We have also advanced our swingarm strategy by building high-order plasmonic structures, which broadens the scope of available configurations and optical responses. Our work outlines a general approach to build complex dynamic plasmonic nanoarchitectures, in which multiple optical elements can be readily transported into designated locations over long distances, resulting in programmed structural reconfigurations with high fidelity. Programmable and dynamic high-order plasmonic structures can find useful applications in different fields, ranging from optical sensing[20] to data storage.[21] In particular, the possibility to translocate optical elements in multiple configurations can be used to explore new approaches to encode information at high density.

**Acknowledgements**

We thank Marion Kelsch for assistance with TEM. TEM images were collected at the Stuttgart Center for Electron Microscopy. We thank the helpful discussions with Dr. L. Xin. This project was financed by the European Union's Horizon 2020 research and innovation program under Grant No. 964995 "DNA-FAIRYLIGHTS", by the Deutsche Forschungsgemeinschaft (DFG, German Research Foundation) - 448727036, and by the Baden-Württemberg Stiftung (Internationale Spitzenforschung, BWST-ISF2020-19).

**Keywords:** DNA origami • Gold nanoparticles • Chiral plasmonics • Circular dichroism spectroscopy • Dynamic nanosystems

[1] a) R. D. Vale, *Cell* **2003**, *112*, 467-480; b) M. G. van den Heuvel, C. Dekker, *Science* **2007**, *317*, 333-336; c) R. H. Fillingame, *Science* **1999**, *286*, 1687-1688; d) G. Woehlke, M. Schliwa, *Nat. Rev. Mol. Cell. Biol.* **2000**, *1*, 50-58; e) F. J. Nedelec, T. Surrey, A. C. Maggs, S. Leibler, *Nature* **1997**, *389*, 305-308.

[2] K. J. Verhey, T. A. Rapoport, *Trends Biochem. Sci.* **2001**, *26*, 545-550.

[3] a) P. W. Rothemund, *Nature* **2006**, *440*, 297-302; b) S. M. Douglas, H. Dietz, T. Liedl, B. Hogberg, F. Graf, W. M. Shih, *Nature* **2009**, *459*, 414-418; c) H. Dietz, S. M. Douglas, W. M. Shih, *Science* **2009**, *325*, 725-730; d) S. M. Douglas, A. H. Marblestone, S. Teerapittayanon, A. Vazquez, G. M. Church, W. M. Shih, *Nucleic Acids Res.* **2009**, *37*, 5001-5006; e) Y. Ke, S. M. Douglas, M. Liu, J. Sharma, A. Cheng, A. Leung, Y. Liu, W. M. Shih, H. Yan, *J. Am. Chem. Soc.* **2009**, *131*, 15903-15908; f) K. Pan, D. N. Kim, F. Zhang, M. R. Adendorff, H. Yan, M. Bathe, *Nat. Commun.* **2014**, *5*, 5578; g) M. R. Jones, N. C. Seeman, C. A. Mirkin, *Science* **2015**, *347*, 1260901.

[4] a) Y. Chen, M. S. Wang, C. D. Mao, *Angew Chem Int Edit* **2004**, *43*, 3554-3557; b) P. Yin, H. Yan, X. G. Daniell, A. J. Turberfield, J. H. Reif, *Angew Chem Int Edit* **2004**, *43*, 4906-4911; c) T. Omabegho, R. Sha, N. C. Seeman, *Science* **2009**, *324*, 67-71; d) K. Lund, A. J. Manzo, N. Dabby, N. Michelotti, A. Johnson-Buck, J. Nangreave, S. Taylor, R. J. Pei, M. N. Stojanovic, N. G. Walter, E. Winfree, H. Yan, *Nature* **2010**, *465*, 206-210; e) M. J. Urban, C. Zhou, X. Y. Duan, N. Liu, *Nano Letters* **2015**, *15*, 8392-8396; f) C. Zhou, X. Y. Duan, N. Liu, *Nature Communications* **2015**, *6*, 8102.

[5] a) P. Ketterer, E. M. Willner, H. Dietz, *Sci. Adv.* **2016**, *2*, e1501209; b) T. Tomaru, Y. Suzuki, I. Kawamata, S. M. Nomura, S. Murata, *Chem. Commun. (Camb)* **2017**, *53*, 7716-7719; c) E. Kopperger, J. List, S. Madhira, F. Rothfischer, D. C. Lamb, F. C. Simmel, *Science* **2018**, *359*, 296-300; d) S. Lauback, K. R. Mattioli, A. E. Marras, M. Armstrong, T. P. Rudibaugh, R. Sooryakumar, C. E. Castro, *Nat. Commun.* **2018**, *9*, 1446; e) Y. Yang, S. Zhang, S. Yao, R. Pan, K. Hidaka, T. Emura, C. Fan, H. Sugiyama, Y. Xu, M. Endo, X. Qian, *Chemistry* **2019**, *25*, 5158-5162; f) Y. Ahmadi, A. L. Nord, A. J. Wilson, C. Hutter, F. Schroeder, M. Beeby, I. Barisic, *Small* **2020**, *16*, e2001855; g) L. Xin, X. Duan, N. Liu, *Nat. Commun.* **2021**, *12*, 3207; h) E. Bertosin, C. M. Maffeo, T. Drexler, M. N. Honemann, A. Aksimentiev, H. Dietz, *Nat. Commun.* **2021**, *12*, 7138; i) A. K. Pumm, W. Engelen, E. Kopperger, J. Isensee, M. Vogt, V. Kozina, M. Kube, M. N. Honemann, E. Bertosin, M. Langecker, R. Golestanian, F. C. Simmel, H. Dietz, *Nature* **2022**, *607*, 492-498; j) A. Buchl, E. Kopperger, M. Vogt, M. Langecker, F. C. Simmel, J. List, *Biophys. J.* **2022**, DOI: 10.1016/j.bpj.2022.1008.1046; k) A. Peil, L. Xin, S. Both, L. Shen, Y. Ke, T. Weiss, P. Zhan, N. Liu, *ACS Nano* **2022**, *16*, 5284-5291.

[6] a) H. Gu, J. Chao, S. J. Xiao, N. C. Seeman, *Nature* **2010**, *465*, 202-205; b) F. C. Simmel, *Curr. Opin. Biotechnol.* **2012**, *23*, 516-521.

[7] a) J. S. Shin, N. A. Pierce, *J. Am. Chem. Soc.* **2004**, *126*, 10834-10835; b) A. J. Thubagere, W. Li, R. F. Johnson, Z. Chen, S. Doroudi, Y. L. Lee, G. Izatt, S. Wittman, N. Srinivas, D. Woods, E. Winfree, L. Qian, *Science* **2017**, *357*, eaan6558; c) P. Zhan, K. Jahnke, N. Liu, K. Gopfrich, *Nat. Chem.* **2022**, *14*, 958-963; d) R. Ibusuki, T. Morishita, A. Furuta, S. Nakayama, M. Yoshio, H. Kojima, K. Oiwa, K. Furuta, *Science* **2022**, *375*, 1159-1164.

[8] a) A. Kuzyk, R. Jungmann, G. P. Acuna, N. Liu, *ACS Photonics* **2018**, *5*, 1151-1163; b) N. Liu, T. Liedl, *Chem. Rev.* **2018**, *118*, 3032-3053.

[9] Y. Hao, Y. Li, L. Song, Z. Deng, *J. Am. Chem. Soc.* **2021**, *143*, 3065-3069.

[10] V. Amendola, R. Pilot, M. Frasconi, O. M. Marago, M. A. Iati, *J. Phys. Condens. Matter* **2017**, *29*, 203002.

[11] a) K. L. Kelly, E. Coronado, L. L. Zhao, G. C. Schatz, *J Phys Chem B* **2003**, *107*, 668-677; b) E. Prodan, C. Radloff, N. J. Halas, P. Nordlander, *Science* **2003**, *302*, 419-422.

[12] a) A. Kuzyk, R. Schreiber, Z. Fan, G. Pardatscher, E. M. Roller, A. Hogele, F. C. Simmel, A. O. Govorov, T. Liedl, *Nature* **2012**, *483*, 311-314; b) M. K. Nguyen, A. Kuzyk, *ACS Nano* **2019**, *13*, 13615-13619; c) M. J. Urban, C. Shen, X. T. Kong, C. Zhu, A. O. Govorov, Q. Wang, M. Hentschel, N. Liu, *Annu. Rev. Phys. Chem.* **2019**, *70*, 275-299; d) C. Zhou, X. Duan, N. Liu, *Acc. Chem. Res.* **2017**, *50*, 2906-2914.

[13] a) Z. Fan, A. O. Govorov, *Nano Lett* **2010**, *10*, 2580-2587; b) M. Hentschel, M. Schaferling, X. Duan, H. Giessen, N. Liu, *Sci. Adv.* **2017**, *3*, e1602735.

[14] a) T. Funck, F. Nicoli, A. Kuzyk, T. Liedl, *Angew. Chem. Int. Ed. Engl.* **2018**, *57*, 13495-13498; b) J. Dong, M. Wang, Y. Zhou, C. Zhou, Q. Wang, *Angew. Chem. Int. Ed. Engl.* **2020**, *59*, 15038-15042.

[15] a) L. X. Chao Zhou, Xiaoyang Duan, Maximilian J. Urban, and Na Liu, *Nano Lett.* **2018**, *18*, 7395–7399; b) Y. Huang, M. K. Nguyen, A. K. Natarajan, V. H. Nguyen, A. Kuzyk, *ACS Appl. Mater. Interfaces* **2018**, *10*, 44221-44225.

[16] a) A. Kuzyk, M. J. Urban, A. Idili, F. Ricci, N. Liu, *Sci. Adv.* **2017**, *3*, e1602803; b) K. Gopfrich, M. J. Urban, C. Frey, I. Platzman, J. P. Spatz, N. Liu, *Nano Lett.* **2020**, *20*, 1571-1577.

[17] Funck T., Liedl T., Bae W., *Appl. Sci.* **2019**, *9*, 3006.

[18] L. Xin, C. Zhou, X. Duan, N. Liu, *Nat. Commun.* **2019**, *10*, 5394.

[19] a) J. Fu, Y. R. Yang, A. Johnson-Buck, M. Liu, Y. Liu, N. G. Walter, N. W. Woodbury, H. Yan, *Nat. Nanotechnol.* **2014**, *9*, 531-536; b) Y. Chen, G. Ke, Y. Ma, Z. Zhu, M. Liu, Y. Liu, H. Yan, C. J. Yang, *J. Am. Chem. Soc.* **2018**, *140*, 8990-8996.

[20] M. Dass, F. N. Gur, K. Kolataj, M. J. Urban, T. Liedl, *J. Phys. Chem. C Nanomater Interfaces* **2021**, *125*, 5969-5981.

[21] a) H. Ditlbacher, B. Lamprecht, A. Leitner, F. R. Aussenegg, *Opt. Lett.* **2000**, *25*, 563-565; b) A. Doricchi, C. M. Platnich, A. Gimpel, F. Horn, M. Earle, G. Lanzavecchia, A. L. Cortajarena, L. M. Liz-Marzan, N. Liu, R. Heckel, R. N. Grass, R. Krahne, U. F. Keyser, D. Garoli, *ACS Nano* **2022**, DOI: 10.1021/acsnano.2c06748.

**Transformable plasmonic helix**

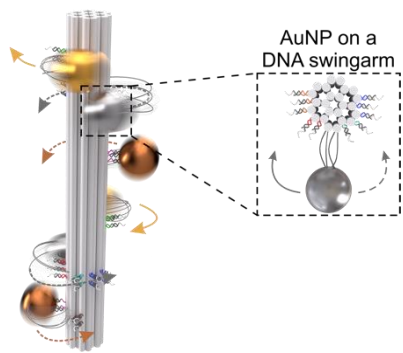

We demonstrate a transformable plasmonic helix, in which multiple AuNPs can programmable move around a shared DNA origami shaft. To enable the translocations of multiple AuNPs in large leaps without detachment, the AuNPs are guided by DNA swingarms extended from the origami to target positions through sequence-specific DNA interactions.